\documentclass{llncs}

\usepackage{graphicx}

\usepackage[latin1]{inputenc} 
\newdimen\proofrulebreadth \proofrulebreadth=.05em
\newdimen\proofdotseparation \proofdotseparation=1.25ex
\newdimen\proofrulebaseline \proofrulebaseline=2ex
\newcount\proofdotnumber \proofdotnumber=3
\let\then\relax
\def\hfi{\hskip0pt plus.0001fil}
\mathchardef\squigto="3A3B
\newif\ifinsideprooftree\insideprooftreefalse
\newif\ifonleftofproofrule\onleftofproofrulefalse
\newif\ifproofdots\proofdotsfalse
\newif\ifdoubleproof\doubleprooffalse
\let\wereinproofbit\relax
\newdimen\shortenproofleft
\newdimen\shortenproofright
\newdimen\proofbelowshift
\newbox\proofabove
\newbox\proofbelow
\newbox\proofrulename
\def\shiftproofbelow{\let\next\relax\afterassignment\setshiftproofbelow\dimen0 }
\def\shiftproofbelowneg{\def\next{\multiply\dimen0 by-1 }%
\afterassignment\setshiftproofbelow\dimen0 }
\def\setshiftproofbelow{\next\proofbelowshift=\dimen0 }
\def\setproofrulebreadth{\proofrulebreadth}

\def\prooftree{
\ifnum  \lastpenalty=1
\then   \unpenalty
\else   \onleftofproofrulefalse
\fi
\ifonleftofproofrule
\else   \ifinsideprooftree
        \then   \hskip.5em plus1fil
        \fi
\fi
\bgroup
\setbox\proofbelow=\hbox{}\setbox\proofrulename=\hbox{}%
\let\justifies\proofover\let\leadsto\proofoverdots\let\Justifies\proofoverdbl
\let\using\proofusing\let\[\prooftree
\ifinsideprooftree\let\]\endprooftree\fi
\proofdotsfalse\doubleprooffalse
\let\thickness\setproofrulebreadth
\let\shiftright\shiftproofbelow \let\shift\shiftproofbelow
\let\shiftleft\shiftproofbelowneg
\let\ifwasinsideprooftree\ifinsideprooftree
\insideprooftreetrue
\setbox\proofabove=\hbox\bgroup$\displaystyle 
\let\wereinproofbit\prooftree
\shortenproofleft=0pt \shortenproofright=0pt \proofbelowshift=0pt
\onleftofproofruletrue\penalty1
}

\def\eproofbit{
\ifx    \wereinproofbit\prooftree
\then   \ifcase \lastpenalty
        \then   \shortenproofright=0pt  
        \or     \unpenalty\hfil         
        \or     \unpenalty\unskip       
        \else   \shortenproofright=0pt  
        \fi
\fi
\global\dimen0=\shortenproofleft
\global\dimen1=\shortenproofright
\global\dimen2=\proofrulebreadth
\global\dimen3=\proofbelowshift
\global\dimen4=\proofdotseparation
\global\count255=\proofdotnumber
$\egroup  
\shortenproofleft=\dimen0
\shortenproofright=\dimen1
\proofrulebreadth=\dimen2
\proofbelowshift=\dimen3
\proofdotseparation=\dimen4
\proofdotnumber=\count255
}

\def\proofover{
\eproofbit 
\setbox\proofbelow=\hbox\bgroup 
\let\wereinproofbit\proofover
$\displaystyle
}%
\def\proofoverdbl{
\eproofbit 
\doubleprooftrue
\setbox\proofbelow=\hbox\bgroup 
\let\wereinproofbit\proofoverdbl
$\displaystyle
}%
\def\proofoverdots{
\eproofbit 
\proofdotstrue
\setbox\proofbelow=\hbox\bgroup 
\let\wereinproofbit\proofoverdots
$\displaystyle
}%
\def\proofusing{
\eproofbit 
\setbox\proofrulename=\hbox\bgroup 
\let\wereinproofbit\proofusing
\kern0.3em$
}

\def\endprooftree{
\eproofbit 
  \dimen5 =0pt
\dimen0=\wd\proofabove \advance\dimen0-\shortenproofleft
\advance\dimen0-\shortenproofright
\dimen1=.5\dimen0 \advance\dimen1-.5\wd\proofbelow
\dimen4=\dimen1
\advance\dimen1\proofbelowshift \advance\dimen4-\proofbelowshift
\ifdim  \dimen1<0pt
\then   \advance\shortenproofleft\dimen1
        \advance\dimen0-\dimen1
        \dimen1=0pt
        \ifdim  \shortenproofleft<0pt
        \then   \setbox\proofabove=\hbox{%
                        \kern-\shortenproofleft\unhbox\proofabove}%
                \shortenproofleft=0pt
        \fi
\fi
\ifdim  \dimen4<0pt
\then   \advance\shortenproofright\dimen4
        \advance\dimen0-\dimen4
        \dimen4=0pt
\fi
\ifdim  \shortenproofright<\wd\proofrulename
\then   \shortenproofright=\wd\proofrulename
\fi
\dimen2=\shortenproofleft \advance\dimen2 by\dimen1
\dimen3=\shortenproofright\advance\dimen3 by\dimen4
\ifproofdots
\then
        \dimen6=\shortenproofleft \advance\dimen6 .5\dimen0
        \setbox1=\vbox to\proofdotseparation{\vss\hbox{$\cdot$}\vss}%
        \setbox0=\hbox{%
                \advance\dimen6-.5\wd1
                \kern\dimen6
                $\vcenter to\proofdotnumber\proofdotseparation
                        {\leaders\box1\vfill}$%
                \unhbox\proofrulename}%
\else   \dimen6=\fontdimen22\the\textfont2 
        \dimen7=\dimen6
        \advance\dimen6by.5\proofrulebreadth
        \advance\dimen7by-.5\proofrulebreadth
        \setbox0=\hbox{%
                \kern\shortenproofleft
                \ifdoubleproof
                \then   \hbox to\dimen0{%
                        $\mathsurround0pt\mathord=\mkern-6mu%
                        \cleaders\hbox{$\mkern-2mu=\mkern-2mu$}\hfill
                        \mkern-6mu\mathord=$}%
                \else   \vrule height\dimen6 depth-\dimen7 width\dimen0
                \fi
                \unhbox\proofrulename}%
        \ht0=\dimen6 \dp0=-\dimen7
\fi
\let\doll\relax
\ifwasinsideprooftree
\then   \let\VBOX\vbox
\else   \ifmmode\else$\let\doll=$\fi
        \let\VBOX\vcenter
\fi
\VBOX   {\baselineskip\proofrulebaseline \lineskip.2ex
        \expandafter\lineskiplimit\ifproofdots0ex\else-0.6ex\fi
        \hbox   spread\dimen5   {\hfi\unhbox\proofabove\hfi}%
        \hbox{\box0}%
        \hbox   {\kern\dimen2 \box\proofbelow}}\doll%
\global\dimen2=\dimen2
\global\dimen3=\dimen3
\egroup 
\ifonleftofproofrule
\then   \shortenproofleft=\dimen2
\fi
\shortenproofright=\dimen3
\onleftofproofrulefalse
\ifinsideprooftree
\then   \hskip.5em plus 1fil \penalty2
\fi
}

\usepackage {bsymb}
\usepackage{fancyhdr,lastpage,color}
\usepackage{url}
  
\newcommand{\ebtojml}{Event\-B\-2\-Jml}
\newcommand{\ebtodafny}{Event\-B\-2\-Dafny}
\newcommand{\jml}{JML}
\newcommand{\eb}{Event-B}

\newcommand{\ebkeyw}[1]{\textsf{#1}}
\newcommand{\ebtag}[1]{\textsf{\bf{#1}}}

\def \bsl {\symbol{92}}

\def \BTOJMLN {\textsf{EB2Jml}}
\newcommand{\BTOJML}[1]{\BTOJMLN\textsf{(}\ensuremath{{#1}}\textsf{)}}
\def \PREDN {\textsf{Pred}}
\newcommand{\PRED}[1]{\PREDN\textsf{(}\ensuremath{{#1}}\textsf{)}}
\def \TON {\textsf{TypeOf}}
\newcommand{\TO}[1]{\TON\textsf{(}\ensuremath{{#1}}\textsf{)}}

\def \MODN {\textsf{Mod}}
\newcommand{\MOD}[1]{\MODN\textsf{(}\ensuremath{{#1}}\textsf{)}}

\begin{document}

\title{A Machine-Checked Proof for a Translation of Event-B Machines
  to JML}

\author{N\'{e}stor Cata\~{n}o{1} \and 
  Camilo Rueda{2} \and
  Tim Wahls{3}
}

\institute{
  EAFIT Unviersity \\
  \email{nestor.catano@gmail.com}
  \and 
  Camilo Rueda
  Pontificia Universidad Javeriana \\
  \email{crueda@cic.puj.edu.co}
  \and
  Tim Wahls 
  Dickinson College \\
  \email{wahlst@dickinson.edu}
}

\maketitle

\begin{abstract}
We present a machine-checked soundness proof of a translation of \eb\
to the Java Modeling Language (\jml). The translation is based on an
operator \BTOJMLN\ that maps \eb\ events to \jml\ method
specifications, and deterministic and non-deterministic assignments to
\jml\ method post-conditions. This translation has previously been
implemented as the \ebtojml\ tool. We adopted a \emph{taking our own
medicine} approach in the formalisation of our proof so that \eb\ as
well as \jml\ are formalised in \eb\ and the proof is discharged with
the Rodin platform. Hence, for any \eb\ substitution (whether an event
or an assignment) and for the \jml\ method specification obtained by
applying \ebtojml\ to the substitution, we prove that the semantics of
the JML method specification is \emph{simulated} by the semantics of
the substitution. Therefore, the \jml\ specification obtained as
translation from the \eb\ substitution is a \emph{refinement} of the
substitution. Our proof includes invariants and the standard \eb\
initialising event, but it does not include full machines or \eb\
\emph{contexts}. We assume that the semantics of \jml\ and \eb\
operate both on the same initial and final states, and we justify our
assumption.

\end{abstract}

\keywords{\eb, \jml, \ebtojml, Formal Proof, Translation, Rodin}

\section{Introduction}
\label{sec:intro}
In an earlier work~\cite{EventB2Jml2013}, we have proposed a design
methodology that combines the use of design-by-contract techniques
with \jml\ \cite{RobbyChalin2009,DBC:JML:06,DBC:Meyer} and
correctness-by-construction techniques with \eb\ \cite{Abrial2010}. In
the proposed methodology, software development starts with a model in
\eb\ which is proven to adhere to a set of safety properties; the \eb\
model is then \emph{refined}, e.g. more details are added to the
model, and the \ebtojml\ tool\footnote{Available at
  \texttt{http://poporo.uma.pt/\-$\sim$ncatano/\-Projects/\-favas/\-EventB2JML.html}}
is used to translate the refined \eb\ model to a \jml\ specification. The
\ebtojml\ tool has been used to develop several
Android~\cite{Android,AndroidDev} applications, including a
social-event planner~\cite{SocialPlanner} and a car racing
game~\cite{TheCarModel}. The translation from \eb\ to \jml\ is based
on an operator \BTOJMLN\ that maps \eb\ events to \jml\ method
specifications.  Several of the rules that define \BTOJMLN\ are
presented in Section~\ref{sec:translation}. 

This paper presents a
machine-checked soundness proof of the translation encoded by
\BTOJMLN. We have adopted a \emph{taking our own medicine} approach in
the formalisation of the proof so that \eb\ as well as \jml\ are
formalised in \eb\ and the proof is discharged in Rodin
\cite{rodinSources}. Our formalisation rests on a \emph{shallow}
embedding \cite{GordonHOL,GordonLogics} of the semantics of \eb\ and
\jml\ in the logic of the \eb\ language. A shallow embedding approach
(as opposed to a deep embedding) facilitates conducting and
discharging a proof, but does not permit proofs of meta-properties
about the language one is modelling. Hence, for any \eb\ substitution
(whether an event or an assignment) and for the \jml\ method
specification obtained by applying \ebtojml\ to the substitution, we
prove that the semantics of the JML method specification is
\emph{simulated} by the semantics of the substitution. In this
respect, the semantics of the \jml\ specification is a
\emph{refinement} of the semantics of the \eb\ model in that any
behaviour in \jml\ is matched by a behaviour in \eb. The soundness
proof considers \eb\ machine invariants, machine variables, events
and, in particular, the \eb\ {initialising} event, but not machines or
machine contexts. To facilitate the proof we make a few assumptions
(e.g. that the \jml\ and \eb\ models operate on the same initial and
final states), and we justify the soundness of our assumptions. The
proof is structured as a collection of contexts in Rodin with
translation and semantic rules implemented as axioms. We used the ProB
model-checker \cite{prob} to find a valuation for our axiomatic
definitions so as to enhance our confidence that no
inconsistencies have been introduced.

The contributions of this paper are the following: $(i.)$ we present a
machine-checked soundness proof for the key rules defining the
\BTOJMLN\ operator. These rules are the core of the implementation of
the \ebtojml\ tool. $(ii.)$ we generalise our proof approach so that
it can be extended to soundness proofs of translations of other formal
languages.

\section{Related Work}
\label{sec:related}

Defining a shallow or a deep embedding of one language in another
is not a new
idea~\cite{GordonHOL,GordonLogics}. Indeed, J. Bowen and M. Gordon
in~\cite{GordonZAndHOL} propose a shallow embedding for
{Z}~\cite{WoodcockBookOnZ} in {HOL}. Catherine Dubois~\textit{et
  al.}~propose a deep embedding of B in the logic of
Coq~\cite{DuboisBinCoq07,Dubois:Deep:2011} for which B proof rules are
formalized and proved correct. 
Jean Paul Bodeveix, Mamoun Filali, and C{\'e}sar
Mu{\~n}oz ~\cite{CMunoz99} generalise the substitution mechanism of
the B method in Coq and PVS as a shallow embedding.
In this current work, we consider a shallow embedding of \eb\ (the successor
of B) into \eb\ itself, which allows us to use Rodin in proving the 
soundness of our translation.

Some other efforts relate to the automated verification of B and \eb\
proof obligations. David D{\'e}harbe presents an approach to translate
a particular subclass of B and \eb\ proof obligations into the input
language of SMT-solvers~\cite{deharbe:SMT}. We have
implemented the \ebtodafny\ Rodin plug-in~\cite{EventB2Dafny} which
translates \eb\ proof obligations into Dafny, one of the front-end
specification languages of the Boogie 2 tool~\cite{boogie-leino-06}.

Other works relate to the translation of formal specification (and
modelling) languages, and to the implementation of tools automating
these translations. In an earlier work
\cite{conference:B2Jml:12}, the authors define a translation from
\emph{classical} B to \jml\ implemented in the ABTools suite
\cite{Boulanger2003}. 

M\'ery and Singh \cite{Mery:2011} define a translation tool
(implemented as a Rodin plug-in) that automatically translates \eb\
machines into several different languages: C, C++, Java and
C\#. Wright \cite{Wright09} defines a B2C extension to the Rodin
platform that translates \eb\ models into C code. However, this work
considers only simple translations of formal concrete machines.
Edmunds and Butler \cite{EB10,EB11} present a tasking extension for
\eb\ that generates code for concurrent programs (targeting
multitasking, embedded and real-time systems).  The main issue with
these tools is that the user has to provide a final (or at least an
advanced) refinement of the system so that it can be directly
translated to code. And, they are limited to integer and boolean
types, so event guards cannot include relations or sets.

Jin and Yang \cite{jin2008} outline an approach for translating VDM-SL
to \jml.  Their reasons for doing so are similar to our motivation for
translating \eb\ to \jml\ -- they view
VDM-SL as a better language for modelling at an abstract level (much
the way that we view \eb), and \jml\ as a better language for working
closer to an implementation level.
Bouquet et.~al. have defined a translation from JML to
B~\cite{groslambert-jml2b-05} and implemented their approach in the
JML2B tool~\cite{groslambert-jml2b-07}. 
Their motivation is again quite
similar to ours -- they view translation as a way to gain access to more
appropriate tools for the task at hand, which in this case is
verifying the correctness of an abstract model without regard to code.
JML verification tools are primarily concerned with verifying the
correctness of code with respect to specifications, while B has much
stronger tool support for verifying models.

\section{Preliminaries}
\subsection{The \eb\ Method}
\eb\ models are composed of \emph{machines} and
\emph{contexts}. Machines contain the dynamic parts of a model
(e.g. variables, invariants, events). Contexts contain the static part
of a model (e.g. carrier sets, constants). Three basic relationships
are used to structure a model. A machine \emph{sees} a context and can
\emph{refine} refine another machine, and a context can \emph{extend}
another context. A partial example of a social networking model
(adapted from the B model in \cite{conference:matelas:2010}) is
depicted in Figure~\ref{fig:ebnet} (abstract and first refinement
machines). Both machines see a context $ctx1$ (not shown) that defines
carrier sets \ebkeyw{PERSON} (the set of all possible people in the
network) and \ebkeyw{CONTENTS} (the set of all possible images, text,
... in the network). The abstract machine declares variables $persons$
(the set of people actually in the network), $contents$ (the set of
content actually in the network), $owner$ (a total surjection mapping
each content item to its owner), and $pages$ (a total relation
indicating which content items are visible to which people). Invariant
\ebtag{inv5} ensures that each content item is visible to its owner.
\eb\ provides notations $\rel$ for a
relation, $\tsur$ for a total surjective function, and $\strel$ for a
total surjective relation.

\begin{figure*}
{\small
\[
\begin{array}{c@{\hspace*{15pt}}c}
\begin{array}{l}
\ebkeyw{machine}~abstract~\ebkeyw{sees}~ctx1\\
~\ebkeyw{variables}~persons~contents~owner~pages \\
~\ebkeyw{invariant}\\
~~\ebtag{inv1 } persons \subseteq \ebkeyw{PERSON}\\
~~\ebtag{inv2 } contents \subseteq \ebkeyw{CONTENTS}\\
~~\ebtag{inv3 } owner \in contents \tsur persons\\
~~\ebtag{inv4 } pages \in contents \strel persons\\
~~\ebtag{inv5 } owner \subseteq pages\\
~\ebkeyw{events}\\
~~initialisation \\
~~~\ebkeyw{begin} \\ 
~~~~\ebtag{act1 }persons := \emptyset \\
~~~~\ebtag{act2 }contents := \emptyset \\
~~~~\ebtag{act3 }owner := \emptyset \\
~~~~\ebtag{act4 }pages := \emptyset \\
~~~\ebkeyw{end} \\
~~create\_account \\
~~~\ebkeyw{any}~p1~c1~\ebkeyw{where}\\
~~~~\ebtag{grd1 }p1 \in \ebkeyw{PERSON} \setminus persons\\
~~~~\ebtag{grd2 }c1 \in  \ebkeyw{CONTENTS} \setminus contents\\
~~~\ebkeyw{then}\\
~~~~\ebtag{act1 }contents := contents \bunion \{c1\}\\
~~~~\ebtag{act2 }persons := persons \bunion \{p1\}\\
~~~~\ebtag{act3 }owner := owner \bunion \{c1 \mapsto p1\}\\
~~~~\ebtag{act4 }pages := pages \bunion \{c1 \mapsto p1\}\\
~~\ebkeyw{end} \\
~~edit\_owned \\
~~~\ebkeyw{any}~c1~p1~newc~\ebkeyw{where}\\
~~~~\ebtag{grd1 }c1 \in contents\\
~~~~\ebtag{grd2 }p1 \in persons\\
~~~~\ebtag{grd3 }owner(c1) = p1\\
~~~~\ebtag{grd4 }newc \in \ebkeyw{CONTENTS} \setminus contents\\
~~~\ebkeyw{then}\\
~~~~\ebtag{act1 }contents := (contents \setminus \{c1\}) \\
~~~~~~~~~~~~~~~~~~~~~~~~\bunion \{newc\}\\
~~~~\ebtag{act2 }pages := (\{c1\}) \domsub pages)\\
~~~~~~~~~~~~~~~~~~~~\bunion (\{newc\} \cprod pages[\{c1\}])\\
~~~~\ebtag{act3 }owner := (\{c1\}) \domsub owner)\\
~~~~~~~~~~~~~~~~~~~~~\bunion (\{newc \mapsto p1\}\\
~~\ebkeyw{end} \\
\ebkeyw{end}
\end{array}
&
\begin{array}{l}
\ebkeyw{machine}~ref1\_permissions\\
 ~~~\ebkeyw{refines}~abstract~~\ebkeyw{sees}~ctx1\\
~\ebkeyw{variables}~persons~contents~owner~pages\\
~~~~~~~~~~~~viewp~editp \\
~\ebkeyw{invariant}\\
~~\ebtag{invr1 } viewp \in contents \rel persons\\
~~\ebtag{invr2 } editp \in contents \rel persons\\
~~\ebtag{invr3 } owner \subseteq viewp\\
~~\ebtag{invr4 } owner \subseteq editp\\
~~\ebtag{invr5 } editp \subseteq viewp\\
~~\ebtag{invr6 } pages \subseteq viewp\\
~\ebkeyw{events}\\
~~initialisation \\
~~~\ebkeyw{begin} \\ 
~~~~\ebtag{act1 }persons := \emptyset \\
~~~~\ebtag{act2 }contents := \emptyset \\
~~~~\ebtag{act3 }owner := \emptyset \\
~~~~\ebtag{act4 }pages := \emptyset \\
~~~~\ebtag{actr1 }viewp := \emptyset \\
~~~~\ebtag{actr2 }editp := \emptyset \\
~~~\ebkeyw{end} \\
~~create\_account~\ebkeyw{extends}~create\_account\\
~~~\ebkeyw{then}\\
~~~~\ebtag{actr1 }viewp := viewp \bunion \{c1 \mapsto p1\}\\
~~~~\ebtag{actr2 }editp := editp \bunion \{c1 \mapsto p1\}\\
~~\ebkeyw{end} \\
~~edit\_owned~\ebkeyw{extends}~edit\_owned\\
~~~\ebkeyw{then}\\
~~~~\ebtag{actr1 }viewp := (\{c1\} \domsub viewp) \\
~~~~~~~~~~~~~~~~~~~~~~\bunion (\{newc\} \cprod viewp[\{c1\}])\\
~~~~\ebtag{actr2 }editp := (\{c1\} \domsub editp) \\
~~~~~~~~~~~~~~~~~~~~~\bunion (\{newc \mapsto p1\}\\
~~\ebkeyw{end} \\
\ebkeyw{end}
\end{array}
\end{array}
\]
}
\caption{Part of an \eb\ machine for social networking
  (left: abstract machine. right: first refinement).}
\label{fig:ebnet}
\end{figure*}

The $initialisation$ event ensures that all of these sets, functions
and relations are initially empty. The symbol $\emptyset$ represents
the empty set. The abstract machine further defines the ``standard''
event $create\_\-account$ that adds a new person and associated
initial content to the network. \eb\ events can be executed/triggered
when their guards (the part after the \ebkeyw{where}) are true. Hence,
the $create\_\-account$ event can execute whenever there is at least
one person and at least one content item that has not yet been added.
The meaning of an event is the meaning of the actions in its body (the
part after the \ebkeyw{then}).  After the $create\_\-account$ event
executes, the person $p1$ and content $c1$ are added to the network,
$p1$ owns $c1$, and $c1$ is visible to $p1$.

The $edit\_\-owned$ event replaces a content item $c1$ with a new
content item $newc$ with the same ownership and visibility. The symbol
$\setminus$ represents set difference, $\mapsto$ a pair of elements,
$\domsub$ domain subtraction, which removes from a relation all the
pairs with a first element in some set, and $\cprod$
cross-product.  The expression $pages[\{c1\}]$ evaluates the relation 
$pages$ on the
set $\{c1\}$. 

The first refinement of this machine (right Figure
\ref{fig:ebnet}) introduces permissions. It adds
variables $viewp$ and $editp$ which respectively track which people
have permission to view and edit each content item (actions
\ebtag{actr1} and \ebtag{actr2}). The invariants further specify
actions stating that that the owner always has permission to view and
edit an item, that a person must have view permission on an item in
order to have edit permission, and that an item is not visible to a
person who does not have view permission on the item. The abstract
event $create\_account$ is extended so that the owner has edit and view
permissions on the item, and $edit\_\-owned$ is extended so that the new
content item has the same edit and view permissions as the item it is
replacing.

\subsection{\jml}
\label{sec:jml}
\jml\ \cite{Chalin-etal2007,JML:ExpRep:05} is a model-based language
designed for specifying the interfaces of Java classes. \jml\
specifications are typically embedded directly into Java class
implementations using special comment markers \texttt{/*@}
\texttt{...}  \texttt{*/} or \texttt{//@}.  Specifications include
various forms of invariants and pre- and post-conditions for
methods. Mathematical types that are heavily used in other model based
specification languages (sets, sequences, relations and functions) are
provided in \jml\ as classes.

 \begin{figure}
{\small
\begin{verbatim}
import poporo.models.JML.*;
import org.jmlspecs.models.JMLEqualsEqualsPair; 

public abstract class ref1_permissions {
/*@ public model BSet<Integer> CONTENTS; */
/*@ public model BSet<Integer> PERSON; */

/*@ public model BSet<Integer> contents;
    public model BRelation<Integer,Integer> editp;
    public model BRelation<Integer,Integer> owner;
    public model BRelation<Integer,Integer> pages;
    public model BSet<Integer> persons;
    public model BRelation<Integer,Integer> viewp; */
/*@ public invariant
      persons.isSubset(PERSON) && contents.isSubset(CONTENTS)
   && owner.isaFunction() && owner.domain().equals(contents) 
   && owner.range().equals(persons) 
   && pages.domain().equals(contents) && pages.range().equals(persons)
   && owner.isSubset(pages)
   && viewp.domain().isSubset(contents) && viewp.range().isSubset(persons) 
   && editp.domain().isSubset(contents) && editp.range().isSubset(persons)
   && owner.isSubset(viewp) && owner.isSubset(editp) 
   && editp.isSubset(viewp) && pages.isSubset(viewp); */

/*@ initially 
      persons.isEmpty() && contents.isEmpty() && owner.isEmpty()
   && pages.isEmpty() && viewp.isEmpty() && editp.isEmpty() ; */ 


/*@ assignable \nothing;
    ensures \result <==> 
     (\exists Integer c1; (\exists Integer p1; (\exists Integer newc;
       (contents.has(c1) && persons.has(p1) && owner.apply(c1) == p1  
    && CONTENTS.difference(contents).has(newc))))); */
public abstract boolean guard_edit_owned();

/*@ requires guard_edit_owned();
    assignable contents, pages, owner, viewp, editp;
    ensures (\exists Integer c1; (\exists Integer p1; (\exists Integer newc; 
       \old(contents.has(c1) && persons.has(p1) && owner.apply(c1) == p1     
    && CONTENTS.difference(contents).has(newc)) 
    && contents.equals(\old(contents.difference(
         new BSet<Integer>(c1)).union(new BSet<Integer>(newc))))
    && pages.equals(\old(pages.domainSubtraction(
         new BSet<Integer>(c1)).union(Utils.cross(new BSet<Integer>(newc),
                                      pages.image(new BSet<Integer>(c1)))))) 
    && owner.equals(\old(owner.domainSubtraction(new BSet<Integer>(c1)).union(
         new BRelation<Integer,Integer>(new JMLEqualsEqualsPair<Integer,Integer>
                                          (newc,p1)))))
    && viewp.equals(\old(viewp.domainSubtraction(new BSet<Integer>(c1)).union(
         Utils.cross(new BSet<Integer>(newc),viewp.image(new BSet<Integer>(c1))))))
    && editp.equals(\old(editp.domainSubtraction((new BSet<Integer>(c1))).union(
         new BRelation<Integer,Integer>(new JMLEqualsEqualsPair<Integer,Integer>
                                          (newc,p1)))))))); 
also
    requires !guard_edit_owned();
    assignable \nothing;
    ensures true; */
    public abstract void run_edit_owned();
}
\end{verbatim}
}

%
\caption{A partial \jml\ specification of a social networking class.}
\label{fig:netJAVA}
\end{figure}

Figure~\ref{fig:netJAVA} presents a partial output of applying the
\ebtojml\ tool to the first refinement machine of the social
networking system model (right Figure \ref{fig:ebnet}). In \jml,
\texttt{model} fields are specification-only -- they are an
abstraction of the mutable part of an object's state, and need not be
directly present in the implementation.  Classes \texttt{BSet} and
\texttt{BRelation} are built-in to \ebtojml\ and represent
mathematical sets and relations, respectively. \eb\ carrier sets are
represented in \jml\ as a set of integers (as instances of
\texttt{BSet<Integer>}). The \texttt{model} fields are translations of
the variables of the \eb\ model in Figure~\ref{fig:ebnet} and have the
same meanings as those variables.  The \texttt{invariant} states
properties that must hold in every visible system state --
specifically before a public method is triggered and after the method
terminates. This is semantically equivalent to conjoining the
invariant to the pre- and post-conditions of each method
specification.

In this example, the first part of the invariant corresponds to the
invariant of the abstract model (see left Figure \ref{fig:ebnet}). In
the \eb\ model, \texttt{owner} is represented as a function, not as
a relation.  The invariant uses the method \texttt{isaFunction()} to
enforce both that \texttt{owner} is a function and that its domain and
range are \texttt{contents} and \texttt{persons} (respectively),
thus defining \texttt{owner} as a total surjection.

The \texttt{initially} clause (which is implicitly conjoined to the
post-condition of every constructor) specifies that all of the fields
representing \eb\ variables are initially empty. The values set by the
\texttt{initially} clause are used as the initial values for checking
class invariants.

An \eb\ event can be triggered (its actions can be executed) when its
guard is satisfied. This behaviour is modelled by creating two
methods: a ``guard'' method containing the translation of the guard,
and a ``run'' method containing the translation of the event body.  In
the ``guard'' method, \verb+\+\texttt{result} represents the result
returned by a method call. The ``run'' method should only be
executed when the ``guard'' method returns \texttt{true}. In system
states that satisfy the translation of the event guard, executing the
``run'' method must result in a state that satisfies the translation
of the body. In system states that do not satisfy the translation of
the event guard, the corresponding event could not be triggered and so
executing the ``run'' method must have no effect.  This is handled by
translating an event to two \jml\ method specification cases -- one
where the translation of the guard is satisfied, and one where it is
not.  This is expressed in the pre-condition (\texttt{requires} clause
in \jml) of each specification case. The first specification case for
\texttt{run\_\-edit\_\-owned} (in which method
\texttt{guard\_\-edit\_\-owned} returns \texttt{true}) specifies
the effect of replacing a content item in the system.  The
\texttt{assignable} clause is a frame condition specifying what
locations may change from the pre-state to the post-state, e.g. all
fields representing \eb\ variables for this case.  The pre-state is
the state on entry to the method and the post-state is the state on
exit from the method. Two special \texttt{assignable} specification
exist, \verb|assignable| \verb|\nothing|, which specifies that the
method modifies no location, and \verb|assignable| \verb|\everything|,
which specifies that the method can modify any. The post-condition
(\texttt{ensures} in \jml) is expressing that in the state on exit
from the method, 
content item \texttt{c1} is replaced by content item \texttt{newc}
with the same ownership, visibility and permissions.
Expressions within \texttt{\bsl old} are evaluated in the
pre-state, while all other expressions are evaluated in the
post-state.  In the second specification case, no locations are
\texttt{assignable}, and so the post-state must be equal to the
pre-state when \texttt{guard\_edit\_owned} returns \texttt{false}
in the pre-state.  Due to space considerations, the translation 
of the $create\_\-account$ event is not included in Figure~\ref{fig:netJAVA}.

\section{The Translation from Event-B to JML}
\label{sec:translation}
The translation from \eb\ to \jml\ is defined with the aid of an
\BTOJMLN\ operator that translates \eb\ syntax to \jml\ syntax. In the
following, we present the \BTOJMLN\ rules for \eb\ substitutions and
for the translation of a machine that are needed for the soundness
proof in Section~\ref{sec:proof}.
A much more complete description of the translation is presented 
in \cite{EventB2Jml2013}.

Events in \eb\ are translated to two \jml\ methods: a \texttt{guard}
method that determines when the guard of the corresponding event
holds, and a \texttt{run} method that models the execution of the
corresponding event. In Rule \textsf{Any} below, variables bound by an
\ebkeyw{any} construct are existentially quantified in the
translation, as any values for those variables that satisfy the guards
can be chosen. The \PREDN\ operator translates \eb\ predicates and
expressions to \jml, so the \texttt{guard} method returns \texttt{true}
when the translation of the guard is satisfied. 

The \jml\
specification of each \texttt{run} method uses two specification
cases (indicated by keyword \texttt{also} in \jml).
In the first case, the translation of the guard is satisfied
and the post-state of the method must satisfy the translation of the
actions.  In the second case, the translation of the guard is not
satisfied, and the method is not allowed to modify any fields,
ensuring that the post-state is the same as the pre-state. This
matches the semantics of \eb\ -- if the guard of an event is not
satisfied, the event cannot execute and hence cannot modify the system
state.  
The translation of the guard is included in the
post-condition of the first specification case of the \texttt{run} method in
order to bind the variables introduced by the \ebkeyw{any},
as they can be used in the body of
the event. Translation of events uses an additional helper operator
\MODN, which calculates the set of variables assigned by the actions
of an event (the \jml\ \texttt{assignable} clause). 

Note that the effective pre-condition of a \jml\ method with
multiple specification cases is the
disjunction of the pre-conditions of each case -- so that in our
translation, the pre-condition of a \texttt{run} method is always
\texttt{true}. Hence, even though guards are translated as
pre-conditions, no method in the translation result has a
pre-condition. Rather, the translation of the guard determines which
behaviour the method must exhibit.  The semantics of multiple
specification cases in \jml\ is implication -- the post-state
of a method must satisfy the post-condition (\texttt{ensures} clause) of
a specification case only when the 
\texttt{requires} clause of that specification case holds. 

\[
\begin{prooftree} 
\begin{tabular}{ll}
\TO{x} = \texttt{Type} \quad
\PRED{G(s,c,v,x)} = \texttt{G} \\
\MOD{A(s,c,v,x)} = \texttt{M} \quad
\BTOJML{A(s,c,v,x)} = \texttt{A} \\
\end{tabular} \vspace*{0.5em}
\using \textsf{(Any)\label{any:label}}
\justifies
\begin{array}{c}
\begin{array}{c}
\BTOJMLN(\ebkeyw{events}\ evt\:\ebkeyw{any}\;x\;\ebkeyw{where}
\;G(s,c,v,x)\\
\quad\quad\quad\quad\quad\quad\ \ \ \ \ \ \;
  \ebkeyw{then}\:A(s,c,v,x) \:\ebkeyw{end})
  =\\
\end{array}
\\
\begin{array}{l}
\texttt{/*@ assignable \bsl nothing;}\\
\quad\quad \,\texttt{ensures \bsl result <==> }
\texttt{(\bsl exists Type x; G); */}\\
\texttt{public abstract boolean guard\_evt();}\\
\\
\texttt{/*@ requires guard\_evt();}\\
\quad\quad \,\texttt{assignable M;}\\
\quad\quad \,\texttt{ensures (\bsl exists Type x;}\;
  \texttt{\bsl old(G) \texttt{\&\&} A);}\\
\quad \, \texttt{also}\\
\quad\quad \, \texttt{requires !guard\_evt();}\\
\quad\quad \,\texttt{assignable \bsl nothing;}\\
\quad\quad \,\texttt{ensures true; */} \\
\texttt{public abstract void run\_evt();}
\end{array}
\end{array}
\end{prooftree}
\]

The translation of ordinary and non-deterministic assignments via
operator \BTOJMLN\ is presented below.  The symbol $:\!\!|$ represents
non-deterministic assignment. Non-deterministic assignments generalise
deterministic assignments (formed with the aid of $:=$), e.g. $v := v
+ w$ can be expressed as $v :\!|\;v' = v + w$.  If variable $v$ is of
a primitive type, the translation would use \texttt{==} rather than
the \texttt{equals} method.

\[
\begin{prooftree}
\PRED{E(s, c, v)} = \texttt{E}
\using \textsf{(Asg)}
\justifies
\begin{array}{c}
\BTOJML{v\::=E\:} = 
\texttt{v.equals(\bsl old(E))}\\
\end{array}
\end{prooftree}
\]

\[
\begin{prooftree}
\PRED{P(s, c, v, v')} = \texttt{P}
\quad \TO{v} = \texttt{Type}
\using \textsf{(NAsg})
\justifies
\begin{array}{c}
\BTOJML{v:\!|\:P} = \\
\texttt{(\bsl exists Type v'; \bsl old(P) \&\&}\;
 \texttt{v.equals(v'))}
\end{array}
\end{prooftree}
\]

Multiple actions in the body of an event are translated individually and
the results are conjoined.  For example, a pair of actions:
\[
\begin{array}{c}
\ebtag{act1 }x := y\\
\ebtag{act2 }y := x\\
\end{array}
\]

is translated to: \texttt{x == \bsl old(y) \&\& y == \bsl old(x)} for integer
variables $x$ and $y$, which correctly models simultaneous actions
as required by the semantics of \eb.

The \MODN\ operator \emph{collects} the variables assigned by \eb\
actions.  The cases of \MODN\ for assignments are shown below. For the
body of an event, \MODN\ is calculated by unioning the variables
assigned by all contained actions.
 \[
 \begin{array}{c@{\hspace*{20pt}}c}
 \begin{array}{l}
 \MOD{v\::=E\:} = \{v\}
 \end{array}
 &
 \begin{array}{l}
 \MOD{v:\!|\:P} = \{v\}
 \end{array}
 \end{array}
 \]

 Rule \textsf{Inv} translates a \jml\ invariant to an \eb\ invariant using
 the \PREDN\ operator.

\[
\begin{prooftree}
\PRED{I(s, c, v)} = \texttt{I}
\using \textsf{(Inv)}
\justifies
\begin{array}{c}
\BTOJML{\ebkeyw{invariants}\;I(s, c, v)} = \\
\texttt{//@ public invariant I;}
\end{array}
\end{prooftree}
\]

 Rule \textsf{M} below translates a machine $M$ that sees a context
 $C$. We assume that all \eb\ proof obligations are discharged
 (e.g. in Rodin) before the machine is translated to JML, so that
 proof obligations and closely related \eb\ constructs (namely,
 \ebkeyw{witness}es and \ebkeyw{variant}s) need not be considered in
 the translation and in the soundness proof. A \ebkeyw{witness}
 contains the value of a disappearing abstract event variable, and a
 \ebkeyw{variant} is an expression that should be decreased by all
 \ebkeyw{convergent} events. An \eb\ machine is translated to a single
 \texttt{abstract} \jml\ annotated Java class, which can be extended
 by a subclass that implements the abstract methods. 
 The translation of the context $C$ is incorporated into the
 translation of machines that see the context.

\[
\begin{tabular}{cc}
$
\begin{prooftree}
\begin{array}{l}
\BTOJML{\ebkeyw{sets}\;s} = \texttt{S}\\
\BTOJML{\ebkeyw{constants}\;c} = \texttt{C}\\
\BTOJML{\ebkeyw{axioms}\;X(s, c)} = \texttt{X}\\
\BTOJML{\ebkeyw{theorems}\;T(s, c)} = \texttt{T}\\
\BTOJML{\ebkeyw{variables}\;v} = \texttt{V}\\
\BTOJML{\ebkeyw{invariants}\;I(s, c, v)} = \texttt{I}\\
\BTOJML{\ebkeyw{events}\: e} = \texttt{E}
\end{array}
\using \textsf{(M)}
\justifies
\begin{array}{l}
\BTOJMLN(\ebkeyw{machine}\;M\;\ebkeyw{sees}\;C \\
\quad\quad\quad\quad\quad \ebkeyw{variables}\;v \\
\quad\quad\quad\quad\quad \ebkeyw{invariants}\;I(s,c,v) \\
\quad\quad\quad\quad\quad \ebkeyw{events}\;e \\
\quad\quad\quad\quad\ebkeyw{end}) =\\
\texttt{public abstract class}\; M\; \texttt{\{}\\
\quad\quad 
\texttt{S} \quad
\texttt{C} \quad
\texttt{X} \quad
\texttt{T} \quad
\texttt{V} \quad
\texttt{I} \quad
\texttt{E}\\
\texttt{\}}
\end{array}
\end{prooftree}
$

&

$
\begin{array}{l}
\ebkeyw{context}\;C \\
\quad\ebkeyw{sets}\;s \\
\quad\ebkeyw{constants}\;c \\
\quad\ebkeyw{axioms}\;X(s, c) \\
\quad\ebkeyw{theorems}\;T(s, c) \\
\ebkeyw{end}
\end{array}
$
\end{tabular}
\]

 The rules defining \BTOJMLN\ for sets, constants, axioms, theorems and 
 variables are not needed for the proof (in the next section), and so are
 not presented here.  The interested reader is invited to 
 consult \cite{EventB2Jml2013}.

\section{The Proof}
\label{sec:proof}

Embedding a language in the logic of a proof assistant consists of
encoding the semantics and syntax of the language in the logic. Two
different ways of formalising languages in logic have been proposed, namely,
\emph{deep} and \emph{shallow}
embedding~\cite{GordonHOL,GordonLogics,HelinDeepShallow}. In the deep
embedding approach, the syntax and the semantics of the language are
formalised in logic as structures, e.g. as data-types. It is thus
possible to prove meta-theoretical properties of the language, but
proofs are usually cumbersome. In the shallow embedding approach, the
language is embedded as types or logical predicates. Proofs become
simpler, although one cannot then prove meta-theoretical properties of
the language itself.

Our proof of the soundness of the translation from \eb\ to \jml\ rests
as a shallow embedding of \eb\ and \jml\ in the logic of the \eb\
language. The proof allows for deterministic and non-deterministic
assignments, events (including the initialising event)
and machine invariants.  Contexts are not considered. Our
embedding abstracts away the translation of \eb\ predicates, and thus
the correctness of the semantics of the whole translation assumes the
correctness of the translation of these predicates. We abstract
machine variables and events, that is, we consider machines to be
composed of a single variable and a (non-initialising) single
event. Our proof of soundness can be extended to consider a set of
variables and a set of events, but we have not done so
here. Therefore, the ability of \eb\ to non-deterministically trigger
enabled events is not part of the soundness proof (or of the
translation presented in Section~\ref{sec:translation}). In this
respect, we prove that a \jml\ specification is a refinement of an
\eb\ model and this proof is approached in a per-event basis just as
refinement proof obligations are generated for \eb\ in tools like
Rodin \cite{RodinHandBook}. Notice that our soundness proof need not
include a proof of absence of deadlocks (the machine invariant entails
the disjunction of the guards of the events) since we assume that all
the proof obligations of the \eb\ model are discharged before it is
translated to \jml.

Roughly speaking, our soundness proof ensures that any state
transition step of the \jml\ semantics of the translation of some \eb\
construct into \jml\ can be \emph{simulated} by a state transition
step of the \eb\ semantics of that construct. All steps in the proof
are modelled in \eb\ and implemented in Rodin~\cite{rodin}, a platform
that provides support for writing and verifying \eb\
models~\cite{rodinSources}. The soundness condition just described is
stated as a theorem and proved interactively in Rodin. The whole proof
structure is represented by a collection of contexts in Rodin, with
translation rules and semantics implemented as axioms. 

Substitutions are of two forms as described in
Figure~\ref{fig:eb:subs}. The substitution on the left
non-deterministically selects a value that satisfies the predicate $P$
and assigns this value to $v$. The symbol ``$:\!|$'' stands for
non-deterministic assignment. The predicate $P$ is a
\emph{before-after} predicate that depends on the value of the machine
variables before ($v$) and after ($v'$) the assignment. The
substitution in right Figure~\ref{fig:eb:subs} (an event) is
parameterised by an event variable $x$. The event may only be executed
(triggered) when the guard $G$ holds. If so, the implementation of the
event may select appropriate values for $v$, $v'$ and $x$ that make
the before-after predicate $Q$ true.

In formalising and conducting the proof of soundness of the
translation from \eb\ to \jml, we assume the following:

\begin{enumerate}
\item Translating an \eb\ predicate produces a \jml\ predicate whose
  semantics is the same as that of the predicate it was translated from. 
\item Translating an \eb\ machine invariant produces a \jml\ 
  invariant whose semantics is the same as of the invariant it was
  translated from.  This is a logical consequence of the previous assumption
  and Rule \textsf{Inv} as presented in Section~\ref{sec:translation}.
%
\item States are defined in the same way in the semantics of both \eb\ and
 \jml\ -- as partial functions from identifiers to values.  We are abstracting
 the differences between values in \eb\ and \jml\ by assuming that all values
 are drawn from a single set $Value$.
\item \eb\ machines are composed of a single machine variable, an
  initialising event, and a single standard event. Machines do not
  \emph{see} any context.
\end{enumerate}

Additionally, we consider a simplified version of \eb\
substitutions. Rule \textsf{NAsg} in Section~\ref{sec:translation}
operates on the substitution in left Figure~\ref{fig:eb:subs}, and
the substitution in right is a simplified version of the one on
which Rule~\textsf{Any} in Section~\ref{sec:translation} operates,
namely, the body of the event in right Figure~\ref{fig:eb:subs}
consists of a single event action.

Our soundness proof proceeds via the following sequence of steps:

\begin{enumerate}
\item expressing \eb\ and \jml\ constructs as types in \eb.
\item  implementing the \ebtojml\ translation rules as type transducer rules.
\item defining a semantics of \eb\ types as state
  transducers.
\item defining a semantics of \jml\ types as state transducers.
\item proving that the semantics of the \jml\ translation of \eb\ constructs
 is simulated by the \eb\ semantics of those constructs.
\end{enumerate}

\begin{figure}[tdp]
\begin{center}
 \[
\begin{array}{ll}
\begin{array}{|l|}
  \hline
  \mathbf{begin}\\
  ~~v :\!|\;P(v,v')\\
  \mathbf{end}\\
  \hline
\end{array}
\begin{array}{l}
~~~~~\\
\end{array} &
\begin{array}{|l|}
\hline
\mathbf{any}\\
~~x\\
\mathbf{where}\\
~~ G(v,x)\\
\mathbf{then}\\
~~v :\!|\;Q(v,v',x)\\
\mathbf{end}\\
\hline
\end{array}
\end{array}
\]
\end{center}
\caption{\eb\ substitutions.}
\label{fig:eb:subs}
\end{figure}

\subsection{The Types for \eb\ and \jml}
\label{subsec:types}
For \eb\ we define types $BPredType$ and $BSubsType$, representing the
type of a predicate and the type of a substitution (a
non-deterministic assignment or an event), together with \emph{type
  constructors} $BPred$ for predicates, $Assg$ for assignments, and
$Any$ for events as below.
Additionally, we define a distinguished initialisation event constructor $BInit$
that produces a substitution. To provide a uniform presentation, we
assume that all predicates in \eb\ involve three identifiers, hence
the arity of the constructor $BPred$. Thus, predicates $P$ and $G$ in
Figure \ref{fig:eb:subs} will actually be represented by
$BPred(v,v',v')$ and $BPred(v,x,x)$, respectively. Variable $v'$ is
the after-value of $v$. The substitution in left
Figure~\ref{fig:eb:subs} is written $Assg(v,v',BPred(v,v',v'))$, where
$BPred(v,v',v')$ models the before-after predicate $P$. This
non-deterministic substitution picks a value that satisfies $P$ and
assigns it to $v$. In general, $BPred$ represents predicates involving
three parameters as follows: the first parameter is the machine
variable, the second is the after-value of that same variable, and the
third is the local event variable. If the third parameter is missing,
as is the case in the substitution in left Figure~\ref{fig:eb:subs},
then we take the third parameter as the second one; if the second one
is missing, as in the case of an event guard, then we choose the
second parameter to be equal to the third one.

\[
\begin{array}{|l|} \hline
  BPred \in Id \times Id \times Id \tfun BPredType \\
  Assg \in Id \times Id \times BPredType \tfun BSubsType \\
  Any \in Id \times BPredType \times Id \times Id \times BPredType \tfun BSubsType \\
  BInit \in Id \times Id \times BPredType \tfun BSubsType \\ \hline
\end{array}
\]

We use \emph{carrier types} to abstract away unnecessary details in
the \eb\ to \jml\ translation. Types defined for \eb\ predicates,
substitutions, and machines are the sets $BPredicate$, $BSubs$, and
$BMachine$ respectively. 

\[
\begin{array}{|l|}
  \hline
\ebkeyw{sets} \\
~~BPredicate\; BSubs\; BMachine \\
\ebkeyw{invariants} \\
~~BPredType \subseteq BPredicate \\
~~BSubsType \subseteq BSubs\\
~~MachineType \subseteq BMachine \\ \hline
\end{array}
\]
 
Types for \jml\ constructs are defined in a similar way as for \eb,
but many more components are involved, e.g. we define $JmlNothing$ and
$JmlEverything$ representing \jml's \texttt{assignable} specifications 
\verb|\nothing|, the empty set, and \verb|\everything|, the set of all
the program identifiers.

\[
\begin{array}{|l|} \hline
JmlNothing \subseteq Id \\
JmlNothing = \emptyset \\
JmlEverything = Id \\ \hline
\end{array}
\]

We consider \jml\ predicate
definitions involving,

 \begin{itemize}
 \item equality of identifiers, as in \verb|x=y|
 \item existentially quantified predicates \verb|(\exists T x;| \verb|...)| 
 \item \jml\ boolean operators \verb|&&| (logical and), \verb+||+ (logical or)
 \item constant predicates \verb|true|, \verb|false|
 \item \jml\ predicates with identifier values taken from the pre-state, as
   in $\mathtt{\backslash{}old(P)}$
 \item \jml\ predicates with identifier values taken from the post-state
   (the default)
 \end{itemize}

 These types appear in the translation to \jml\ of \eb\
 constructs. Hence, \eb\ predicates are translated to semantically
 equivalent \jml\ predicates, non-deterministic assignments are
 translated to existentially quantified predicates (rule \textsf{NAsg}
 in Section~\ref{sec:translation}), and events to \jml\ specified
 methods (rule \textsf{Any} in Section~\ref{sec:translation}). JML
 predicates appear in the \texttt{requires} (pre-condition) and
 \texttt{ensures} (post-condition) parts of \jml-specified
 methods. For each predicate listed we define a type (a subset of
 $JmlPredicate$), e.g. $JmlExistsType$, $JmlAndType$, etc. and a type
 constructor. Type constructors are shown below. Notice that, as for
 \eb\ predicates, a simple \jml\ predicate represented by the
 $JmlPred$ constructor involves exactly three identifiers.
 $JmlMeth$ constructs a \jml\ method specification from a \jml\ normal
 ($JmlNormal$) and exceptional ($JmlExceptional$) behaviour
 specifications. Normal and exceptional specifications are composed of
 a pre-condition, a frame condition (the set of variables that may be
 modified), and a normal and exceptional post-condition,
 respectively. $JmlOld$ matches the \jml\ \verb+\old+ operator, and
 $JmlBecomes$ relates the value of machine variables in the pre-state
 with the value in the post-state.

\[
\begin{array}{|l|}
  \hline
  JmlExists \in Id \times JmlPredicate \tfun JmlExistsType\\
  JmlAnd \in JmlPredicate \times JmlPredicate \tfun JmlAndType\\
  JmlTrue \in  JmlTrueType\\
  JmlFalse \in  JmlFalseType\\
  JmlBecomes \in  Id \times Id  \tfun JmlBecomesType\\
  JmlNot \in JmlPredicate \tfun JmlNotType \\
  JmlOld \in JmlPredicate \tfun JmlOldType\\
  JmlPred \in Id \times Id\times Id \tfun   JmlPredType\\
  JmlMeth \in JmlNormalType \times JmlExceptionalType \tfun JmlMethType\\
  JmlNormal \in JmlPredicate \times \pow(Id) \times JmlPredicate \tfun JmlNormalType \\
  JmlExceptional \in JmlPredicate \times \pow(Id) \times JmlPredicate \tfun JmlExceptionalType\\
\hline 
\end{array} 
\]


\subsection{The \eb\ to \jml\ Translation Rules}
The translation rules of Section~\ref{sec:translation} are modelled as
total functions transforming \eb\ types into \jml\ types and
predicates. We consider five functions: one for translating \eb\
predicates, one for non-deterministic assignments, one for standard
events, one for initialising events, and one for machine
invariants. These functions are shown below. An \eb\ predicate is
transformed into a \jml\ predicate, a non-deterministic assignment is
transformed into a \jml\ predicate that is used within a \jml\
postcondition specification, a standard event becomes a \jml\ method
specification, an initialising event becomes the postcondition of a
\jml\ (Java) constructor, and an Event-B machine invariant is
translated to a \jml\ class invariant.

\[
\begin{array}{|l|} \hline
BPred2Jml \in BPredicate \tfun JmlPredicate\\
Assg2Jml \in BSubs \tfun JmlPredicate \\
Any2Jml \in BSubs \tfun JmlMethType \\ 
BInit2Jml \in BPredicate \tfun JmlPredicate \\
BInv2Jml \in BInvType \tfun JmlPredicate \\
\hline
\end{array}
\]

The translation rules for \eb\ substitutions (and other \eb\
constructors) are expressed as axioms. Simple substitutions are
presented in left Figure~\ref{fig:eb:subs}, where $P$ is a
before-after predicate involving variables $v$ and $v'$. 
Since this substitution is not
guarded (meaning there is no condition required for the substitution
to be executed), its translation amounts to the translation of $P$, as
shown by Rule \textsf{NAsg} in Section~\ref{sec:translation}. This is
expressed by the following axiom, where $BPred(v,v',v')$ models $P$,
$JmlOld$ evaluates $P$ in the pre-state, and $JmlBecomes$ expresses
how the value of the machine variable changes from the prestate to the
poststate.

\[
\begin{array}{|l|} \hline
\forall v,v'.~(v\in Id \wedge v'\in Id \Rightarrow \\
~~~~Assg2Jml(Assg(v,v',BPred(v,v',v'))) \\
~~~~~~~~= \\
~~~~JmlExists(v',JmlAnd(JmlOld(BPred2Jml(BPred(v,v',v'))),\\
~~~~~~~~~~~~~~~~~~~~~~~~JmlBecomes(v,v')~))) \\ \hline
\end{array}
\]

Rule \textsf{Any} in Page \pageref{any:label} formalises the
translation of the event shown in Figure~\ref{fig:eb:subs}. The event
implements a guarded substitution. The translation of the event
includes a normal and an exceptional behaviour specification cases. If
the guard of the event holds, the method may modify the machine
variable $v$, and the post-condition of the method correctly
implements the effect produced by the event ($JmlNormal$). If the
guard of the event does not hold ($JmlExceptional$), the method is not
allowed to modify any variable, and it produces no effect.


\[
\begin{array}{|l|} \hline
\forall v,v',x.~(v\in Id \wedge v'\in Id \wedge x\in Id \Rightarrow \\
~~Any2Jml(Any(x,BPred(v,x,x),v,v',BPred(v,v',x))) \\
~~~= \\
~~JmlMeth(\\
~~~~JmlNormal(\\
~~~~~~JmlExists(x,BPred2Jml(BPred(v,x,x))), \\
~~~~~~\{v\},\\
~~~~~~JmlExists(x,JmlAnd(JmlOld(BPred2Jml(BPred(v,x,x))),\\
~~~~~~~~~~~~~~~~~~~~~~~~~~JmlExists(v',JmlAnd(JmlOld(BPred2Jml(BPred(v,v',x))),\\
~~~~~~~~~~~~~~~~~~~~~~~~~~~~~~~~~~~~~~~~~~~~~~~~JmlBecomes(v,v'))) )) ) ,\\
~~~~JmlExceptional(JmlNot(JmlExists(x,BPred2Jml(BPred(v,x,x)))),\\
~~~~~~JmlNothing,\\
~~~~~~JmlTrue )\\
~~)\\
)\\
\hline
\end{array}
\]

\subsection{Semantics}
\eb\ and \jml\ semantics are defined as state transition systems.
Given any two states $a,b$ and an \eb\ substitution $S$, we show that
if the pair $(a,b)$ belongs to the state transition of the \jml\
semantics of the translation of the type of $S$ ($MachineType$,
$AssgType$, $AnyType$, etc) into \jml, then $(a,b)$ also belongs to
the \eb\ semantics of the type of $S$. We assume that state
definitions in the semantics of \eb\ and in \jml\ are both the
same. For each construct in \eb\ and \jml, we define a total function
that holds whenever a given pair of states represents a valid
transition for that construct. Similarly, the semantics of \eb\ and
\jml\ predicates is given by a predicate in the semantic domain that
holds in a state whenever the B or \jml\ predicate is true in that
state. The types involved in the definition of the \eb\ semantic
functions are shown next. States are partial functions that map
identifiers to values.  Function $BPredSem$ provides a semantics for
\eb\ predicates, $BAssgSem$ for non-deterministic assignments,
$BAnySem$ for events, and $BInitSem$ for the initialising
event. $BPredSem$ returns all the states in which the predicate holds,
the other types return a pair of states, the pre- and the
post-states. The semantics of $BAssgSem$, $BAnySem$, and $BInitSem$
depend on the machine invariant, represented as an element of type
$BPredicate$.

\[
\begin{array}{|l|} \hline
  State = (Id \pfun Value)\\ 
  BPredSem \in BPredicate \tfun \pow(State) \\
  BAssgSem \in BSubs \times BPredicate \tfun  \pow(State \leftrightarrow State)\\
  BAnySem \in BSubs \times BPredicate \tfun \pow(State \leftrightarrow State) \\
  BInitSem \in BSubs \times BPredicate \tfun \pow(State \leftrightarrow State) \\ 
\hline
\end{array}
\]

A transition from state $a$ to state $b$ for the execution of the
substitution in left Figure~\ref{fig:eb:subs} exists if and only if 
the machine invariant holds in $a$ and $b$, and there exists some
post-state value $y$ for $v'$ such that the predicate $P(v,y)$ holds
in state $a\ovl\{(v',y)\}$, and the valuation of identifier $v$ in state
$b$ is equal to $y$. The symbol $\ovl$ stands for relation overriding
so that $a\ovl\{(v',y)\}$ denotes the state $a$ modified so that $a(v')$
is equal to $y$. This is asserted in the following axiom.


\[
\begin{array}{|l|} \hline
\forall v,v',a,b.~ (~v\in dom(a) \wedge v'\in Id \wedge v'\not\in dom(a) \wedge v'\neq v \wedge a\in State \wedge b\in State\;\Rightarrow\\
~~(~(a,b) \in  BAssgSem(Assg(v,v',BPred(v,v',v')),BInv(v,v,v)) \\
~~~~~~~~~~\Leftrightarrow\\
~~~~~(~a \in  BPredSem(BInv(v,v,v))\;\wedge \\
~~~~~~~b \in BPredSem(BInv(v,v,v))\;\wedge \\
~~~~~~~~\exists y.~(~y \in  Value\;\wedge \\
~~~~~~~~~~~~~~~(a\ovl \{(v',y)\}) \in  BPredSem(BPred(v,v',v'))\;\wedge \\
~~~~~~~~~~~~~~~b = (a\ovl \{(v,y)\})\\
~~~~~~~~~~)\\
~~~~~) \\
~~) \\
) \\
\hline
\end{array}
\]

For the event in right Figure~\ref{fig:eb:subs}, a transition from
state $a$ to state $b$ is valid if only if $(1)$ the machine invariant
holds in $a$ and $b$, $(2)$ the predicate $G(v,x)$ holds in a state
$a$ that maps some value $y$ to variable $x$, and $(3)$ there exists
some value $z$ for $v'$ such that $P(v,v',x)$ holds in state $a$
modified so that $v'$ has value $z$ and $x$ has value $y$; otherwise
$(2)$ if the predicate $G(v,x)$ does not hold in state $a$, then
states $a$ and $b$ are the same. $BInitSem$ (which is not shown here)
is defined similarly to $BAssgSem$.


\[
\begin{array}{|l|} \hline
\forall v,v',x,a,b.~(~v\in dom(a)\,\wedge v\in dom(b)\,\wedge v'\in Id\,\wedge v'\not\in dom(a)\,\wedge v'\not\in dom(b)\,\wedge\\
~~~x\in Id\,\wedge x\not\in dom(a)\,\wedge v'\neq v\,\wedge x\neq v\,\wedge a\in State\,\wedge b\in State\Rightarrow \\
~~(~(a,b) \in BAnySem(Any(x,BPred(v,x,x),v,v',BPred(v,v',x)),BInv(v,v,v))\\
~~~~~~\Leftrightarrow \\
~~~~(~(~a\in BPredSem(BInv(v,v,v))\,\wedge \\
~~~~~~~~b\in BPredSem(BInv(v,v,v))\,\wedge \\
~~~~~~~~(~\exists y.~(~y \in Value\,\wedge (a\ovl \{(x,y)\}) \in BPredSem(BPred(v,x,x))\,\wedge \\
~~~~~~~~~~~~\exists z.~(~z\in Value\,\wedge (a\ovl \{(x,y)\} \ovl \{(v',z)\}) \in BPredSem(BPred(v,v',x))\,\wedge \\
~~~~~~~~~~~~~~~~~~~~b = a\ovl \{(x,y)\}\ovl \{(v,z)\}~)\\ 
~~~~~~~~~~~~~)\\
~~~~~~~~~)\\
~~~~~~~)\;\vee\\
~~~~~~~(\neg \exists y.~(~y \in Value\,\wedge (a\ovl \{(x,y)\}) \in BPredSem(BPred(v,x,x))\,\wedge (a = b))) \\
~~~~~)\\
~~)\\
)\\
\hline
\end{array}
\]

As with the \eb\ semantics, we define various total functions for
\jml\ constructs, e.g. predicates and method specifications, that hold
for a pair of states when they represent a valid transition from the
first state (the pre-state) to the second (the post-state). These
functions and their types are presented below. $JmlPredSem$ provides
the semantics of a predicate in \jml, $JmlNormalSem$ and
$JmlExcSem$ the semantics of the normal and exceptional
termination of a method, and $JmlMethSem$ the semantics of a \jml\
method specification. These last three constructs are parameterised by
a \jml\ invariant of type $JmlPredicate$.

\[
\begin{array}{|l|} \hline
JmlPredSem \in JmlPredicate \tfun (State \leftrightarrow State)\\
JmlNormalSem \in JmlNormalType \times JmlPredicate \tfun (State \leftrightarrow State)\\
JmlExcSem \in JmlExceptionalType \times JmlPredicate \tfun (State \leftrightarrow State)\\
JmlMethSem \in JmlMethType \times JmlPredicate \tfun (State \leftrightarrow State)\\
\hline
\end{array}
\]

The axiom below provides the semantics of $JmlNormalSem$ for method
pre-condition $req$, method post-condition $ens$, frame-condition
$asg$, and \jml\ invariant $inv$. The symbol $\domres$ stands for
domain restriction, hence the state ``$(Id\,\backslash asg) \domres
a$'' only maps elements in the domain ``$Id\,\backslash asg$''. The
pair of states $(a,b)$ are in the semantics of the normal behaviour
specification of a method if $a$ and $b$ each adhere to the \jml\
invariant, the method precondition - evaluated in the pre-state -
implies the method post-condition, and $a$ is equal to $b$ except
possibly for the variables in the set $asg$. The semantics of
$JmlExcSem$ is defined similarly, and is not shown here.

\[
\begin{array}{|l|} \hline
\forall~req,asg,ens,inv,a,b.~(~a\in State\,\wedge b\in State\,\wedge \\
~~~~req\in JmlPredicate\,\wedge asg\subseteq Id\,\wedge ens\in JmlPredicate\,\wedge inv\in JmlPredicate \Rightarrow \\
~~(~(a,b)\in JmlNormalSem(JmlNormal(req,asg,ens),inv) \\
~~~~~~~~\Leftrightarrow \\
~~~~(~(a,a)\in JmlPredSem(inv) \wedge (b,b)\in JmlPredSem(inv)\, \wedge \\
~~~~~~(a,a)\in JmlPredSem(req) \Rightarrow (a,b)\in JmlPredSem(ens)\, \wedge \\
~~~~~~(Id\,\backslash asg) \domres a = (Id\,\backslash asg) \domres b \\
~~~~) \\
~~) \\
) \\
\hline
\end{array}
\]

The semantics of a \jml\ method specification holds for a pre-state
$a$ and a post-state $b$ if and only if the normal and exceptional
behaviour specifications hold for the same states. This is shown by the
axiom below. 

\[
\begin{array}{|l|} \hline
\forall req1,req2,asg1,asg2,ens1,ens2,inv,a,b.~(~a\in State \wedge b\in State\,\wedge \\
~~~~req1\in JmlPredicate \wedge asg1\subseteq Id \wedge ens1\in JmlPredicate \\
~~~~req2\in JmlPredicate \wedge asg2\subseteq Id \wedge ens2\in JmlPredicate\,\wedge \\
~~~~inv\in JmlPredicate \Rightarrow \\
~~(~(a,b) \in JmlMethSem(JmlMeth(\\
~~~~~~~~~~~~~~~~~~JmlNormal(req1,asg1,ens1), \\
~~~~~~~~~~~~~~~~~~JmlExceptional(req2,asg2,ens2)),inv) \\
~~~~~~~~~~~~~~\Leftrightarrow \\
~~~~(a,b) \in JmlNormalSem(JmlNormal(req1,asg1,ens1),inv)\,\wedge \\
~~~~(a,b) \in JmlExcSem(JmlExceptional(req2,asg2,ens2),inv) \\
~~) \\
) \\
\hline
\end{array}
\]

\jml\ predicates evaluated in a post-state $b$ can refer to predicates
evaluated in a pre-state $a$ using the \jml\ \verb|\old| operator. How
these states relate is best seen in the axiom below.

\[
\begin{array}{|l|} \hline
\forall p,a,b.~(~a\in State \wedge b\in State \wedge p\in JmlPredicate \Rightarrow \\
~(~(a,b) \in JmlPredSem(JmlOld(p)) \\
~~~~~~~\Leftrightarrow \\
~~~(a,a) \in JmlPredSem(p) \\
~) \\
) \\
\hline
\end{array}
\]

The predicate $JmlBecomes$ specifies how the value of a variable can
change from the pre-state to the post-state of a method. $JmlBecomes$
holds for a pair of states $(a,b)$ whenever the post-state state $b$
is the same as pre-state $a$ with identifier $v'$ removed and the
value it had given to identifier $v$. That is, the value of variable
$v'$ in the pre-state is represented by the value of $v$ in the
post-state. The \eb\ symbol $\domsub$ is called domain subtraction,
hence ``$\{v'\}\domsub a$'' removes all the pairs from $a$ whose first
element is $v'$. The symbol $\ovl$ is the symbol for relation
overriding introduced above, hence the state ``$a\ovl \{v,a(v')\}$''
binds $v$ to $a(v')$.

\[
\begin{array}{|l|} \hline
\forall v,v',a,b.~(~a\in State \wedge b\in State \wedge v\in Id \wedge v'\in Id\, \wedge \\
~~~~v\in dom(b) \wedge v'\not\in dom(b) \wedge v'\in dom(a) \wedge v\in dom(a) \wedge v'\neq v \Rightarrow \\
~~(~(a,b)\in JmlPredSem(JmlBecomes(v,v')) \\
~~~~~~\Leftrightarrow \\
~~~~b = \{v'\}\domsub (a\ovl \{(v,a(v'))\}) \\
~~) \\
) \\
\hline
\end{array}
\]





The axiom for existentially quantified predicates is shown below. It
says that $JmlExists(x,p)$ holds for a pre-state $a$ and a post-state
$b$ if and only if $b$ includes $a$, $b$ binds $x$ to some value $y$,
and it does not include any other binding, except for those that might
be created within $p$.

\[
\begin{array}{|l|} \hline
\forall x,p,a,b.~\exists y.~(~a\in State\,\wedge b\in State\,\wedge x\in Id\,\wedge \\
~~~~x\not\in dom(a)\,\wedge y\in Value\,\wedge p\in JmlPredicate\ \Rightarrow \\
~~(~(a,b) \in JmlPredSem(JmlExists(x,p))\\
~~~~~~\Leftrightarrow\\		
~~~(~(a\ovl \{(x,y)\})\subset b\;\wedge\;a = (dom(a) \domres b)\;\wedge \\
~~~~~((a\ovl \{(x,y)\}),b) \in JmlPredSem(p)\\
~~~)\\
)\\
\hline
\end{array}
\]

The semantics of other \jml\ boolean operators follows. The boolean
operator $JmlTrue$ holds for any pair of states and $JmlFalse$ for
none.

\[
\begin{array}{|l|} \hline
JmlPredSem(JmlTrue) = State \times State \\
JmlPredSem(JmlFalse) = \emptyset \\
\hline
\end{array}
\]





\subsection{The Proof Statement}
\label{subsec:pr:stmt}

\begin{figure}[t] 
   \centering
   \includegraphics[width=1.8in]{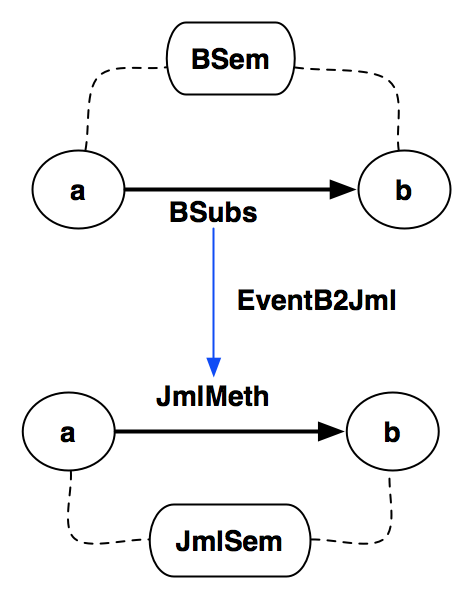} 
   \caption{Proof structure}
   \label{fig:proofStructure}
\end{figure}

The semantics of JML and \eb\ relate as described in Figure 
\ref{fig:proofStructure}.
Hence, for any pair
of states $(a,b)$, and for any \jml\ method obtained via translation from
an \eb\ substitution (non-deterministic assignment or event), if the pair
of states is an element of the semantics of the \jml\ method, then it is
also an element of the semantics of the \eb\ substitution that was translated.
We assume that translation of
\eb\ predicates to \jml\ is correct. The expression of the relation
shown in the figure is achieved through the theorem that follows. The
function $Any2Jml$ produces an element of type $JmlMethType$.


\begin{theorem}
The translation of an event is sound:
\[
\begin{array}{|l|}\hline
\forall x,v,v',a,b.~(~x\in Id\,\wedge v\in Id\,\wedge v'\in Id\,\wedge\, a\in State\,\wedge b\in State\,\wedge \\
~~~v\in dom(a)\,\wedge v\in dom(b)\,\wedge v'\not\in dom(a)\,\wedge v'\not\in dom(b)\,\wedge x\not\in dom(a)\,\wedge \\
~~~v'\neq v\,\wedge x\neq v\,\wedge x\neq v'\Rightarrow \\
~~(~(a,b) \in JmlMethSem(Any2Jml(Any(x,BPred(v,x,x),v,v',BPred(v,v',x))),\\
~~~~~~~~~~~~~~~~~~~~~~~~~~~~~~~~~~~~JmlInv(v,v,v) ) \\
~~~~~~\Rightarrow \\
~~~~(a,b) \in BAnySem(Any(x,BPred(v,x,x),v,v',BPred(v,v',x)),BInv(v,v,v)) \\
~~) \\
)\\
\hline
\end{array}
\]
\end{theorem}

Notice that we state a weak form of semantic correctness in the above
theorem. As previously mentioned, we want to guarantee that any valid
transition of the \jml\ method produced by the translation must also
be a valid transition of the source \eb\ substitution. That is, an
\eb\ substitution must be capable of simulating any valid transition
of its \jml\ method counterpart. The \jml\ translation then
constitutes a sort of ``refinement'' of the \eb\ specification.

The theorem above is discharged under the assumptions below. The
translation of an \eb\ predicate (machine invariant) on some given
variables produces the same predicate (class invariant) as in \jml.

\[
\begin{array}{|l|} \hline
\forall t,u,z.~(~t\in Id \wedge u\in Id \wedge z\in Id\;\Rightarrow\\
~~~~~~~~ BPred2Jml(BPred(t,u,z)) = JmlPred(t,u,z)~) \\ \\
\forall t,u,z,s.~(~t\in Id \wedge u\in Id \wedge z\in Id \wedge s\in State\;\Rightarrow\\
~~~~~~~~ JmlPredSem(JmlPred(t,u,z),s) = BPredSem(BPred(t,u,z),s)~) \\ \\
\forall t,u,z.~(~t\in Id \wedge u\in Id \wedge z\in Id\;\Rightarrow \\
~~~~~~~~ BInv2Jml(BInv(t,u,z)) = JmlInv(t,u,z)~) \\ \\
\forall t,u,z,s.~(~t\in Id \wedge u\in Id \wedge z\in Id \wedge s\in State\;\Rightarrow\\
~~~~~~~~ JmlPredSem(JmlInv(t,u,z),s) = BPredSem(BInv(t,u,z),s)~) \\ \hline
\end{array}
\]

The whole formalization in Rodin consists of 11 machines, 97 axioms,
28 proof obligations (POs) and 1 main theorem. About 60 percent of
the POs were discharged automatically with Rodin, while the remaining
40 percent required manual assistance -- mainly to prove that type
definitions were called with parameters of the correct type. This is
much like discharging TCCs (Type Correctness Conditions) in PVS.

To reduce the chances of introducing inconsistencies, we have
successfully used the ProB model-checker \cite{prob} to find a
valuation that makes all the above axioms true except for the ones
pertaining to methods (including class constructors).
Model-checking of methods timed out. Hence, we stated and
successfully model-checked weaker axioms asserting the
\emph{existence} of states obeying the semantics. This is because our
shallow embedding semantics does not account for the internal
structure of predicates that ProB needs to model check the methods and
constructors, e.g. to model check the relationship between states
obeying a method precondition and states obeying a method
postcondition.

\section{Conclusion}
\label{sec:conclusion}

The ability to transition from \eb\ to \jml\ while developing a
software system has significant practical benefits: an abstract model
of the system can be fully verified in \eb\ and then translated to
\jml\ for implementation.  This reduces the number of proof
obligations that must be discharged by shortening the refinement
chain, and allows programmers who are not familiar with \eb\ notation
and formal refinement techniques to produce a correct system
implementation from the \jml\ specification.  Automating the
translation from \eb\ to \jml\ greatly reduces the overhead associated
with this approach and allows the soundness of the translation to be
proven once, rather than being considered during each development
effort.

The work presented in this paper proves the soundness of our
translation under a precisely documented set of assumptions.  In
particular, we have shown that a \jml\ class specification produced by
the \ebtojml\ tool is a refinement of the \eb\ machine being
translated.  Conducting the proof in Rodin gives a high degree of
confidence in the correctness of the proof itself, allowing us to
focus our attentions on axiomatizing the translation and the semantics
of \eb\ and \jml\ in a concise and effective manner.

Formalising a proof in a tool is often a trade off between the level of
detail the proof should provide and the feasibility of discharging it
in an automated tool.  In our earlier work on the definition of the
\BTOJMLN\ operator~\cite{EventB2Jml2013}, \eb\ types such as sets, functions
and relations are represented using bespoke Java classes (with full \jml\ 
specifications).
The soundness proof presented in Section~\ref{sec:proof} relies on the
correctness of the representation of \eb\ mathematical types by
respective \jml\ types, and, indeed, attempting to represent those in
the logic of \eb\ would greatly increase the complexity of the
proof. An alternative approach would be to use existing \jml\
machinery~\cite{Burdy-etal05,JML:ExpRep:05} to separately verify that
the \jml\ representations do in fact have the same behaviour as the
\eb\ types that they represent.

We have discussed several additional simplifications and assumptions made in
our formal model of \eb\ machines at the beginning of
Section~\ref{sec:proof}.   The assumption that states in \eb\ and \jml\ 
have the same representation is partially addressed in the previous 
paragraph.  Additionally, in Java (and \jml), two references can be aliased
(refer to the same object) in a particular state.  Modelling such states
 requires an extra level of indirection -- typically one partial function 
 is used to map identifiers to locations, and a second to map locations
 to values.
 However, the \jml\ class specifications produced by our translation can not
 be used to create aliases, and so our state definitions need not be capable
 of modelling it.

We assumed that the body of an event is
composed of a single assignment, although, in general, in \eb, multiple
assignments can be executed in parallel using the $||$ operator. In the
specific case of our translation, the right-hand side of an assignment is
always evaluated in the pre-state (due to the use of \texttt{\bsl old} in
the \jml\ translation), so multiple assignments could be composed in any
order in the semantics.  In general, 
a parallel composition of assignments can be translated to sequential
assignments using temporary variables.
That is, the parallel composition $x:= E\;||\;y := F$ can be
expressed as the sequence of assignments $temp := F$ \verb+;+ $x := E$
\verb|;| $y := temp$.  This approach permits translation of parallel
assignments to notations (such as programming languages) that do not provide
a mechanism for evaluating predicates in the pre-state.
Considering \eb\ machines with multiple variables, 
invariants and/or events requires a degree of extra machinery in the
translation and proof, but would not change our fundamental approach or
conclusions.

The five steps enumerated at the beginning of Section~\ref{sec:proof}
outline a general approach for using an automated tool to prove the
soundness of a translation between two formal languages, particularly
when the semantics of the languages are expressed via transition
systems.
Figure~\ref{fig:proofStructure}
in Section~\ref{subsec:pr:stmt} suggests a granularity level for those
transitions in the underlying logic. 

First, the syntactic constructs of both languages are represented in the
logic of the automated tool. Choosing the right representation for
those constructs and the right level of detail will determine the
level of complexity of the proof together with the number of
proof-obligations that are to be discharged (the trade off between a deep
and a shallow embedding). Second, the translation from the source to the
target language is modelled in logic as axioms taking syntactic constructs from
one language to the other. Third and fourth, the type semantics for the
source and the target language are provided, and type constructors to build
elements of those types are defined. Type states are defined
as state transducers. Fifth, a soundness proof is enunciated as
relating state transitions in the target language with state
transitions in the source language. The level of granularity of those
transitions affects the level of complexity of the proof.  To
simplify the proof, it may be useful to make (valid) assumptions about the
representation of states in
the transition system for the source and the target language.
Choosing different representations for the states will
certainly require constructing representations of both
in the logic of the automated tool, and establishing a relationship between
these representations.  This would significantly increase the complexity of
the proof.

It is true that for the approach just described in general (and for 
the proof presented in this paper in particular), what is actually being shown
is the soundness of an axiomatization of the translation with respect to
axiomatizations of the semantics of the source and target languages.
This is a disadvantage of any approach based on shallow embedding -- 
the logical mechanisms required to do the embedding do introduce possibilities
for errors and inconsistencies.  In terms of our translation, the translation
rules and axiomatizations of those rules are given at a similar level of
abstraction and the relationship between them should be immediately apparent.
We do not know of a way to truly validate our axiomatizations of \eb\ and 
\jml\ semantics, but presenting the semantics in such a formal and unambiguous
way does precisely document any assumptions that we have made and otherwise
allows the reader to check the semantics against their own knowledge and 
understanding.


\paragraph{Future Work} We are currently working on a translation from
\jml\ to Java, thus completing the translation from \eb\ down to code. We
will conduct another correctness proof for this new translation. For
Java code generation, we envision a framework in which events,
implemented as Java methods, are non-deterministically invoked by
programming threads when event guards hold. Thus, event guards prevent
updates of the system state. This can be implemented as a subroutine in
which access to a critical section (the event action) is
conditioned by the event guard \cite{HansenMultiP:72}.  
Within this framework, the implementation of standard critical-section
algorithms, e.g. Dekker's \cite{EWD:EWD123pub} or Lamport's bakery
algorithm \cite{Bakery:74}, ensure that any valid sequence of \jml\
method executions (the critical section) correctly simulates the
sequence of event triggering in the \eb\ model that was translated.
No additional correctness proofs need be conducted for this
interaction as properties like mutual exclusion,
absence of deadlocks and starvation for these
algorithms have already been proved in literature.

\bibliographystyle{spmpsci} 
\bibliography{eventb2jml-proof}
\end{document}